\documentstyle[aps,preprint]{revtex}
\tightenlines

\begin{document}

\title{Dual theories for mixed symmetry fields.\\
Spin-two case: (1,1) versus (2,1) Young symmetry type fields.}
\author{H. Casini $^{a}$, R. Montemayor $^b$ and Luis F. Urrutia$^c$}
\address{
$^a$Theoretical Physics, University of Oxford, 1 Keble Road,\\
Oxford OX13NP, United Kingdom \\
$^b$Instituto Balseiro and CAB, Universidad Nacional de Cuyo and CNEA, \\
8400 Bariloche, Argentina\\
$^c$Departamento de F\'\i sica de Altas Energ\'\i as, Instituto
de Ciencias Nucleares \\Universidad Nacional Aut\'onoma de
M\'exico, A.P. 70-543, M\'exico D.F. 04510, M\'exico}
\maketitle
\begin{abstract}
We show that the parent Lagrangian method gives a natural
generalization of the dual theories concept for non $p$-form
fields. Using this generalization we construct here a
three-parameter family of Lagrangians that are dual to the
Fierz-Pauli description of a free massive spin-two system. The
dual field is a three-index tensor $T_{(\mu \nu )\rho }$, which
dinamically belongs to the $(2,1)$ representation of the Lorentz
group. As expected, the massless limit of our Lagrangian, which
is parameter independent, has two propagating degrees of freedom
per space point.
\end{abstract}

\section{Introduction}
In general there is a lot freedom to choose variables for the
description of a physical system. For example, in some cases it
is desirable to have a formulation where some symmetries are
explicit, and this requires the use of a redundant set of
variables to describe the system configuration. In other cases,
as when we perform a canonical quantization, it might be more
convenient to have a minimal, non redundant, set of variables.
The actual proof of the equivalence between different
descriptions is usually a non-trivial task. Equivalent
descriptions of a given physical theory in terms of a different
choice of fields are said to be dual to each other, and the
relation among the fields provides the corresponding duality
transformation \cite{DUAL}.

One of the simplest examples is the scalar-tensor duality. It
corresponds to the equivalence between a free massless scalar
field $\varphi $, with field strength $f_{\mu }=\partial _{\mu
}\varphi $, and a massless antisymmetric field $B_{\mu \nu }$,
the Kalb-Ramond field, with field strength $H_{\mu \nu \sigma
}=\partial _{\mu }B_{\nu \sigma }+\partial _{\nu }B_{\sigma \mu
}+\partial _{\sigma }B_{\mu \nu }$ \cite{FLUND,KR,DSH}. The former
description provides the equation of motion $\partial ^{\mu
}f_{\mu }=0$, together with the Bianchi identities $\partial
^{\mu }\epsilon _{\mu \nu \sigma }{}^{\tau }f_{\tau }=0$. In
turn, the dual description provides the equations of motion
$\partial ^{\mu }H_{\mu \nu \sigma }=0$ and the Bianchi
identities $\partial ^{\mu }\epsilon _{\mu }^{\;\;\nu \sigma \tau
}H_{\nu \sigma \tau }=0$. The duality transformation
\begin{equation}
f_{\mu }\;\;\leftrightarrow \;\;\frac{1}{3}\epsilon _{\mu
}^{\;\;\nu \sigma \tau }H_{\nu \sigma \tau }
\end{equation}
interchanges dynamical equations with Bianchi identities, giving
a full correspondence between both descriptions.

An important predecessor of the modern approach to duality is the
electric-magnetic symmetry $({\vec{E}}+i{\vec{B}})\rightarrow
e^{i\phi }({ \vec{E}}+i{\vec{B}})$ of the free Maxwell equations.
When there are charged sources this symmetry can be maintained by
introducing magnetic monopoles \cite{MILTON}. This transformation
provides a connection between weak and strong couplings via the
Dirac quantization condition. At the level of Yang-Mills theories
with spontaneous symmetry breaking this kind of duality is
expected, due to the existence of topological dyon-type solitons
\cite{OLIVE}. The extension of electromagnetic duality to
$SL(2,Z)$ is usually referred to as S-duality, and plays an
important role in the non-perturbative study of field and string
theories \cite{GOMEZ}.

These basic ideas have been subsequently generalized to arbitrary
forms in arbitrary dimensions. Well known dualities are the ones
between massless $p$ -forms and $(d-p-2)$-forms fields and
between massive $p$ and $(d-p-1)$-forms in $d$ dimensional
space-time\cite{FQUEV1}. These dualities among free fields have
been proved using parent Lagrangians \cite{PL} as well as the
canonical formalism \cite{LOZ}. They can be extended to include
source interactions \cite{QUEV2}.

The above duality among forms can be understood as a relation
between fields in different representations of the Lorentz group.
The origin of this equivalence can be traced using the little
group technique for constructing the representations of the
Poincare group in $d$ dimensions. Given a standard momentum for a
massive or a massless particle, the actual degrees of freedom are
determined by its spin components, which are given by the
irreducible representations of $SO(d-1)$ and $SO(d-2)$
respectively. These are expressed by traceless tensors with a
definite permutation symmetry characterized by the Young
diagrams. For the orthogonal groups $O(n)$, the sum of the
lengths of the first two columns of the Young diagrams is
constrained to be less than or equal to $n$. Two Young diagrams
having their first column of length $l$ and $n-l$ respectively,
with the rest of the diagrams being identical, are called
associated. For the group $SO(n)$ associated Young diagrams
correspond to the same representation \cite{HAM}. The existence
of two different tensorial realizations for one irreducible spin
representation suggests that we can construct two theories with
fields of different tensorial character for the same physical
system. The relation between two equivalent theories expressed in
terms of fields corresponding to associated Young diagrams should
be interpreted as a duality transformation.

In particular, the usual duality among $p$-forms can be
interpreted in this way. A (traceless) $p$-form belongs to a
Young diagram with only one column and $p$ sites, and a
$(d-p-1)$-form belongs to a Young diagram with one column and
$(d-p-1)$ sites. Thus they give the same representation of $
SO(d-1)$ and have the same spin; therefore, if the masses are
equal there is a possibility of constructing alternative theories
for the same physical system. The same can be said regarding the
massless case if we consider $p$ and $(d-p-2)$ forms.

The preceding discussion suggests the possibility of generalizing
the duality transformations among $p$-forms to tensorial fields
with arbitrary Young symmetry types. The simplest generalization
in four dimensions is provided by a massive second-rank symmetric
tensor $h_{\mu \nu }$, with Young symmetry $(1,1)$, and a three
index tensor $T_{\mu \nu \rho }$ with Young symmetry $(2,1)$.
Consistent massless free \cite{FRMIX}, interacting \cite{AG}, and
massive \cite{t0} theories of mixed Young symmetry tensors have
been constructed in the past, but the attempts to prove a dual
relation between these descriptions did not lead to a positive
answer \cite{t0}. Additional interest in this type of theories
arises from the recent formulation of $d=11$ dimensional
supergravity as a gauge theory for the ${\rm osp}(32|1)$
superalgebra. It includes a totally antisymmetric fifth-rank
Lorentz tensor one form $b_{\mu }{}^{abcde}$, whose mixed
symmetry piece does not have any related counterpart in the
standard $\,d=11$ supergravity theory \cite{JZ}.

In this work we show that such dual descriptions can be
constructed. We use a generalization of the parent Lagrangian
method to construct a family of Lagrangians for the massive field
$\,T_{(\mu \nu )\rho }$, which are dual to the standard
Fierz-Pauli Lagrangian for a massive spin two field $h_{\mu \nu
}$. In contrast with the $p$-forms case, we obtain a
multiparameter family of duality transformations. Most notably,
the kinetic part of the dual $\,T$-Lagrangians is unique, and the
parameters appear only in the mass term. The massless limit of
these descriptions is well-behaved in the sense that it has the
correct two propagating degrees of freedom, in contrast with the
absence of degrees of freedom found in previous attempts\cite{t0}.

\section{Massive spin-two irreducible representations}
In the following, we perform a detailed construction for the
massive spin-two fields
in four dimensions. There are two associated Young diagrams,
which correspond to a traceless symmetric rank-two tensor, and to
a traceless rank-three tensor, antisymmetric in two indices, and
satisfying a Jacobi identity. The most usual known description is
given in terms of a symmetric tensor $h_{\mu \nu }=h_{\nu \mu }$,
with the dynamics defined by the Fierz-Pauli Lagrangian
\begin{equation}
{\cal L}=-\partial _{\mu }h^{\mu \nu }\partial _{\alpha }h_{\nu
}^{\alpha }+ \frac{1}{2}\partial _{\alpha }h^{\mu \nu }\partial
^{\alpha }h_{\mu \nu }+\partial _{\mu }h^{\mu \nu }\partial _{\nu
}h_{\alpha }^{\alpha }-\frac{1}{ 2}\partial _{\alpha }h_{\mu
}^{\mu }\partial ^{\alpha }h_{\nu }^{\nu }-\frac{ M^{2}}{2}\left(
h_{\mu \nu }h^{\mu \nu }-h_{\mu }^{\mu }h_{\nu }^{\nu }\right) .
\label{hlag}
\end{equation}
>From the equation of motion it follows that
\begin{equation}
\left( \partial ^{2}+M^{2}\right) h_{\mu \nu }=0,\qquad \partial
_{\mu }h^{\mu \nu }=0,\qquad h_{\mu }{}^{\mu }=0,
\end{equation}
which leads to the five degrees of freedom corresponding to a
massive spin-two irreducible representation.

An alternative description for the free massive spin-two field
corresponding to the associated $(2,1)$ Young diagram has been
already proposed\cite{t0}. It is based on the tensor field
$T_{(\mu \nu )\sigma }$ that satisfies
\begin{equation}
T_{(\mu \nu )\sigma }=-T_{(\nu \mu )\sigma }\;\;\;;\;\;\;T_{(\mu
\nu )\sigma }+T_{(\nu \sigma )\mu }+T_{(\sigma \mu )\nu }=0.
\label{SYMM}
\end{equation}
At this stage $T_{(\mu \nu )\sigma }$ has $20$ independent
components. The proposed Lagrangian is
\begin{equation}
{\cal L}=-\frac{1}{36}\left\{ \left( F_{(\mu \nu \sigma )\tau
}\right) ^{2}-3\left( F_{(\mu \nu \sigma )}{}^{\mu }\right)
^{2}-3M^{2}\left[ \left( T_{(\mu \nu )\sigma }\right)
^{2}-2\left( T_{(\mu \nu )}{}^{\mu }\right) ^{2}\right] \right\}
\label{LAGCURT}
\end{equation}
with the field strength given by
\begin{equation}
F_{(\mu \nu \sigma )\tau }=\partial _{\mu }T_{(\nu \sigma )\tau
}+\partial_{\nu }T_{(\sigma \mu )\tau }+\partial _{\sigma
}T_{(\mu \nu )\tau }. \label{FIELDS}
\end{equation}
In this case the equations of motion imply:
\begin{equation}
\left( \partial ^{2}+M^{2}\right) T_{(\mu \nu )\sigma }=0,\quad
T_{(\mu \nu )}{}^{\mu }=0,\quad \partial ^{\mu }T_{(\mu \nu
)\sigma }=0,\quad \partial ^{\sigma }T_{(\mu \nu )\sigma }=0.
\end{equation}
The first algebraic condition gives four identities. The second
derivative condition is a consequence of the first one plus the
cyclic identity in (\ref {SYMM}). The first derivative condition
includes four identities, $\partial ^{\mu }\partial ^{\nu
}\,T_{\mu \nu \rho }=0$, plus one more, $\partial ^{\mu }\,T_{\mu
\rho }{}^{\rho }$, when we consider independently the zero-trace
condition. This leads to $11$ independent derivative conditions.
The final count produces $5$ independent degrees of freedom, as
appropriate to a massive spin-two field. It is natural to suspect
that both theories could be related by a duality transformation,
but the attempts to construct such a transformation have had no
success \cite{t0}. As we will show below the Lagrangian
(\ref{LAGCURT}) is only one of the possible descriptions for an
irreducible massive spin two field in terms of a $T_{(\mu \nu
)\sigma }$ tensor, and in fact is not a suitable choice to
construct a dual transformation.

\section{Parent Lagrangian for dual theories}
A standard procedure to construct alternative descriptions for a
physical system, which turn out to be related by a duality
transformation, consists in finding a quadratical parent
Lagrangian that contains both types of fields, from which each
theory can be obtained by eliminating either one of them through
the corresponding equations of motion. In this work we show that
a generalization of this method, already successfully applied to
$p$-forms, allows us to construct dual theories for the massive
spin-two representation. Contrary to the approach in \cite{t0},
we start from a tensor $T_{(\mu \nu )\sigma }$ which is only
antisymmetric in the $\mu \nu $ indices, without a priori
satisfying the cyclic identity, together with the standard
symmetric tensor $h_{\mu \nu }$. Eliminating the field $T_{(\mu
\nu )\sigma } $ from the parent Lagrangian we impose the
resulting Lagrangian for $h_{\mu \nu }$ to be the Fierz-Pauli one
describing a massive spin-two system. This provides relations
among the parameters of the model. Once these relations are
implemented in the parent Lagrangian, the elimination of field
$h_{\mu \nu }$ leads to the required dual Lagrangian in terms of
field $T_{(\mu \nu )\sigma }$. The necessary Lagrangian
constraints leading to the $(2,1)$ Young symmetry type of the
field together with the required five independent degrees of
freedom come from the corresponding Euler-Lagrangian equations.

Let us recall the general structure of the parent Lagrangian used
to establish duality between a massive $d-q-1$ form $L$ and a
massive $q$ form $ B$
\begin{equation}
{\cal L}_{P}=\frac{1}{2}L\wedge {}^{*}L+L\wedge
dB+\frac{M^{2}}{2}B\wedge {}^{*}B,  \label{FPL}
\end{equation}
which we will generalize to our case. The most general first
order bilinear Lagrangian for $T_{(\mu \nu )\sigma }$ and $h_{\mu
\nu }$ has seven bilinears in $T$
\begin{eqnarray}
&&T_{(\mu \nu )\sigma }T^{(\mu \nu )\sigma },\quad T_{(\mu \nu
)}^{\nu }T_{\sigma }^{(\mu \sigma )},\quad T_{(\mu \nu )\sigma
}T^{(\mu \sigma )\nu
},  \nonumber \\
\epsilon ^{\mu \nu \alpha \beta }T_{(\mu \nu )\sigma }T_{(\alpha
\beta )}^{\sigma }\,,\quad &&\epsilon ^{\mu \nu \alpha \beta
}T_{(\mu \nu )\sigma }T_{\;\;\alpha )\beta }^{(\sigma }\,,\quad
\epsilon ^{\mu \nu \alpha \beta }T_{(\mu \sigma )\nu
}T_{\;\;\alpha )\beta }^{(\sigma }\,,\quad \epsilon ^{\mu \nu
\alpha \beta }T_{(\mu \nu )\alpha }T_{(\beta \lambda )}^{\lambda
},
\end{eqnarray}
two in $h$,
\begin{equation}
h_{\mu \nu }h^{\mu \nu },\quad h_{\mu }^{\mu }h_{\nu }^{\nu },
\end{equation}
and seven mixing (duality generating) terms which contain both
fields,
\begin{eqnarray}
&&T_{(\mu \nu )\sigma }\partial ^{\mu }h^{\nu \sigma }\,,\quad
T_{(\mu \nu )}^{\;\;\;\;\nu }\partial ^{\mu }h_{\sigma }^{\sigma
}\,,\quad T_{(\mu \nu
)}^{\;\;\;\;\nu }\partial ^{\sigma }h_{\sigma }^{\mu },  \nonumber \\
T_{(\mu \nu )\sigma }\epsilon ^{\mu \nu \alpha \beta }\partial
_{\alpha }h_{\;\beta }^{\sigma },\quad &&T_{(\mu \nu )\sigma
}\epsilon ^{\mu \nu \sigma \beta }\partial _{\alpha }h_{\;\beta
}^{\alpha }\,,\quad T_{(\mu \nu )\sigma }\epsilon ^{\mu \nu
\sigma \beta }\partial _{\beta }h_{\;\alpha }^{\alpha }\,,\quad
T_{(\mu \nu )\sigma }\epsilon ^{\mu \sigma \alpha \beta }\partial
_{\beta }h_{\;\alpha }^{\nu }\,.
\end{eqnarray}

In the following, and for the sake of simplicity, we will explore
in detail only the Lagrangian with one mixing term, $T_{(\mu \nu
)\sigma }\epsilon ^{\mu \nu \alpha \beta }\partial _{\alpha
}h_{\;\beta }^{\sigma }$ , because this is the most natural
generalization of the term $L\wedge d\,B$ in (\ref {FPL}). Thus,
we take our parent Lagrangian to be
\begin{eqnarray}
L &=&a\,T_{(\mu \nu )\sigma }T^{(\mu \nu )\sigma }+b\,T_{(\mu \nu
)}{}^{\nu }T^{(\mu \sigma )}{}_{\sigma }+c\,T_{(\mu \nu )\sigma
}T^{(\mu \sigma )\nu }+d\,\epsilon ^{\mu \nu \alpha \beta
}T_{(\mu \nu )\sigma }T_{(\alpha \beta
)}{}^{\sigma }  \nonumber \\
&&+m\,\epsilon ^{\mu \nu \alpha \beta }T_{(\mu \nu )}{}^{\sigma
}T_{\;\;(\sigma \alpha )\beta }+n\,\epsilon ^{\mu \nu \alpha
\beta }T_{(\mu \sigma )\nu }T^{(\sigma }{}_{\alpha )\beta
}+p\epsilon ^{\mu \nu \alpha
\beta }T_{(\mu \nu )\alpha }T_{(\beta \lambda )}{}^{\lambda }  \nonumber \\
&&+eT_{(\mu \nu )\sigma }\epsilon ^{\mu \nu \alpha \beta
}\partial _{\alpha
}h_{\;\beta }^{\sigma }  \nonumber \\
&&+fh_{\mu \nu }h^{\mu \nu }+kh_{\mu }{}^{\mu }h_{\nu }{}^{\nu }.
\label{plag}
\end{eqnarray}
The equations of motion for $h_{\mu \nu }$ allow us to solve
algebraically this field in terms of $T_{(\mu \nu )\sigma }$
\begin{equation}
h^{\mu \nu }=\frac{e}{4f}\left( \epsilon ^{\alpha \beta \sigma
\nu }\partial _{\sigma }T_{(\alpha \beta )}{}^{\mu }+\epsilon
^{\alpha \beta \sigma \mu }\partial _{\sigma }T_{(\alpha \beta
)}{}^{\nu }\right) -g^{\mu \nu }\frac{ke }{2\left( f+4k\right)
f}\epsilon ^{\alpha \beta \rho \kappa }\partial _{\rho
}T_{(\alpha \beta )\kappa }.  \label{t1}
\end{equation}

In a similar way, from the Euler-Lagrange equation for $T^{(\mu
\nu )\sigma } $ we can algebraically solve this field in terms of
$h^{\mu \nu }$
\begin{equation}
C\,T^{(\kappa \lambda )\sigma }=-e\left( 2m+n-4d\right)
D^{\lambda \sigma \kappa }+\frac{e}{2}\left( 2a+c\right) \epsilon
_{\mu \nu }{}^{\kappa \lambda }D^{\nu \sigma \mu
}-\frac{e}{2}E\,\epsilon _{\mu }{}^{\sigma \kappa \lambda }D^{\mu
}-\frac{e}{2}F\,G^{\sigma \kappa \lambda }\,.  \label{t2}
\end{equation}
where
\begin{eqnarray}
D^{\lambda \sigma \kappa } &\equiv &\partial ^{\lambda }h^{\sigma
\kappa }-\partial ^{\kappa }h^{\sigma \lambda },\qquad D^{\mu
}\equiv \partial _{\beta }h^{\beta \mu }-\partial ^{\mu }h_{\beta
}^{\beta }=D_{\sigma
}^{\;\sigma \mu },  \nonumber \\
G^{\sigma \kappa \lambda } &\equiv &g^{\sigma \kappa }\left(
\partial
_{\alpha }h^{\alpha \lambda }-\partial ^{\lambda }h_{\alpha
}^{\alpha }\right) -g^{\sigma \lambda }\left( \partial _{\alpha
}h^{\alpha \kappa }-\partial ^{\kappa }h_{\alpha }^{\alpha
}\right) =g^{\sigma \kappa }D^{\lambda }-g^{\sigma \lambda
}D^{\kappa }
\end{eqnarray}
and
\begin{eqnarray}
A &=&\left( 4d+m-3p+2n\right) \left[ \left( 4d+m-3p+2n\right)
^{2}+2(2a+c+3b)\left( a-c\right) \right] ^{-1},  \nonumber \\
B &=&-2(2a+c+3b)\left[ \left( 4d+m-3p+2n\right)
^{2}+2(2a+c+3b)\left(
a-c\right) \right] ^{-1},  \nonumber \\
C &=&\left( 2m+n-4d\right) ^{2}+\left( 2a+c\right) ^{2}\,,  \nonumber \\
E &=&\left( 2m+n-4d\right) \left( 2bA+(m+n-p)B\right) +\left(
2a+c\right)
\left( 2(m+n-p)A+cB\right) ,  \nonumber \\
F &=&\left( 2m+n-4d\right) \left( 2(m+n-p)A+cB\right) +\left(
2a+c\right) \left( 2bA+(m+n-p)B\right) \,.
\end{eqnarray}

The parent Lagrangian we are considering contains seven bilinears
in $T$, two bilinears in $h$, and one mixing term, which means we
have started with ten parameters. Nevertheless, only eight
combinations appear in the dual transformations (\ref{t1}) and
(\ref{t2}): $a$, $b$, $c$, $(m+n-p)$, $ (2m+n-4d)$, $e$, $f$, and
$k$. Note that $(4d+m-3p+2n)=3(m+n-p)-(2m+n-4d)$. We further fix
appropriate combinations by using Eq. (\ref{t2}) to rewrite the
parent Lagrangian (\ref{plag}) in terms of $h^{\mu \nu }$ only,
and subsequently demanding it to be the Fierz-Pauli Lagrangian
for a massive spin two field. The result is
\begin{equation}
C=2e^{2}\left( 2a+c\right) ,\,\quad E=2\left( 2a+c\right)
\,,\quad f=-k=- \frac{M^{2}}{2},  \label{equdef}
\end{equation}
The parameters $d$, $m$, $n$ and $p$ only appear through the
combinations $ \left( 2m+n-4d\right) $ and $(m+n-p)$, each of
which can be written in terms of $a$, $b$, $c$ and $e$ using
Eqs. (\ref{equdef}). Considering that the same equations of motion
are obtained up to an overall factor in the Lagrangian, we have
constructed a three-parameter family of parent Lagrangians for a
spin-two field with mass $M$ leading to the Fierz-Pauli
Lagrangian for $ h_{\mu \nu }$.

\section{Spin-two mixed symmetry description}
Now we consider the dual description of the massive spin two
system in terms of the field $T^{(\mu \nu )\rho }$. To this end
we use Eq. (\ref{t1}) and rewrite the parent Lagrangian as
\begin{eqnarray}
L &=&-\frac{e^{2}}{3M^{2}}\left( \left( F_{(\mu \nu \sigma )\tau
}\right) ^{2}-\frac{3}{2}\left( F_{(\mu \nu \sigma )}{}^{\mu
}\right) ^{2}+\frac{1}{2} \left( \epsilon ^{\mu \nu \alpha \beta
}\,\partial _{\beta }T_{(\mu \nu
)\alpha }\right) ^{2}\right)  \nonumber \\
&&+a\,T_{(\mu \nu )\sigma }T^{(\mu \nu )\sigma }+b\,T_{(\mu \nu
)}{}^{\nu }T^{(\mu \sigma )}{}_{\sigma }+c\,T_{(\mu \nu )\sigma
}T^{(\mu \sigma )\nu }+d\,\epsilon ^{\mu \nu \alpha \beta
}\,T_{(\mu \nu )\sigma }T_{(\alpha
\beta )}{}^{\sigma }  \nonumber \\
&&+m\,\epsilon ^{\mu \nu \alpha \beta }\,T_{(\mu \nu )}{}^{\sigma
}T_{(\sigma \alpha )\beta }\ +n\,\epsilon ^{\mu \nu \alpha \beta
}\,T_{(\mu \sigma )\nu }T^{(\sigma }{}_{\alpha )\beta
}+p\,\epsilon ^{\mu \nu \alpha \beta }\,T_{(\mu \nu )\alpha
}T_{(\beta \lambda )}{}^{\lambda }. \label{DUALL}
\end{eqnarray}
The second term of the kinetic part of this Lagrangian has a
coefficient $ \left( -\frac{3}{2}\right) $ instead of the
corresponding $\left( -3\right) $ in the Lagrangian
(\ref{LAGCURT}). This fact was recognized in \cite{t0} as a
potential problem for dualization. However, here we do not impose
the cyclic property (\ref{SYMM}) for $T_{(\mu \nu )\sigma }$ from
the beginning, and this is manifested in the existence of the
third term in the kinetic Lagrangian, which is crucial for
dualization.

It turns out that the trace $T_{\mu }\equiv T_{(\mu \beta
)}{}^{\beta }$ is an auxiliary field in the above Lagrangian.
This is made explicit by defining
\begin{equation}
T_{(\mu \nu )\sigma }=\hat{T}_{(\mu \nu )\sigma
}-\frac{1}{3}\left( g_{\mu \sigma }T_{\nu }-g_{\nu \sigma }T_{\mu
}\right) ,
\end{equation}
where $\hat{T}_{(\mu \nu )\sigma }$ is a traceless field,
$\hat{T}_{(\mu \nu )}{}^{\nu }=0$. Rewriting the Lagrangian
(\ref{DUALL}) in terms of $\,\hat{T} ^{(\chi \psi )\sigma }$ and
$T_{\mu }$ and using the Euler-Lagrangian equation for \ $T_{\mu
}$ we obtain
\begin{equation}
T^{\beta }=\frac{4d+m+2n-3p}{2\left( 2a+3b+c\right) }\epsilon
^{\mu \nu \alpha \beta }\,\hat{T}_{(\mu \nu )\alpha },
\end{equation}
which allows us to algebraically eliminate the trace in the
Lagrangian (\ref {DUALL}). We further introduce the field
strength $\hat{F}_{(\alpha \beta \gamma )\nu }$ appropriate to
$\hat{T}^{(\mu \nu )\sigma }$, defined in (\ref {FIELDS}). In
this way, up to a global factor, (\ref{DUALL}) can be written as
\begin{eqnarray}
L &=&\frac{4}{9}\hat{F}_{(\alpha \beta \gamma )\nu
}\,\hat{F}^{(\alpha \beta \gamma )\nu
}+\frac{2}{3}\hat{F}_{(\alpha \beta \gamma )\nu }\,\hat{F}
^{(\alpha \beta \nu )\gamma }-\hat{F}_{(\alpha \beta \mu
)}{}^{\mu }\,\hat{F}
^{(\alpha \beta \nu )}{}_{\nu }  \nonumber \\
&&+P\,\hat{T}_{(\nu \rho )\sigma }\hat{T}^{(\nu \rho )\sigma
}+Q\,\hat{T} _{(\nu \rho )\sigma }\hat{T}^{(\nu \sigma )\rho
}+R\,\epsilon ^{\mu \nu \alpha \beta }\,\hat{T}_{(\mu \nu )\sigma
}\hat{T}_{(\alpha \beta )}{}^{\sigma }+\lambda ^{\mu
}\hat{T}_{(\mu \beta )}{}^{\beta }, \label{LAGT1}
\end{eqnarray}
where we have introduced the Lagrange multiplier $\lambda ^{\mu
}$ to enforce the traceless condition upon $\hat{T}_{(\mu \sigma
)\nu }$. Here it is
\begin{equation}
P=\frac{2M^{2}}{3e^{2}}\left( \frac{1}{B}-(2a+c)\right)
,\;\;\;Q=-\frac{ 2M^{2}}{3e^{2}}\left( \frac{2}{B}+(2a+c)\right)
, \;\;\; R=\frac{ M^{2}}{2e^{2}}\left( 2m+n-4d\right) .
\end{equation}
The equations (\ref{equdef}) translate in the following relations
between these parameters
\begin{eqnarray}
12\,P^{2}-3\,Q^{2}+16\,R^{2} &=&0,  \nonumber \\
\frac{1}{4}(2P+Q)+4\frac{R^{2}}{(2P+Q)} &=&-M^{2}\text{.}
\label{rell}
\end{eqnarray}
This family of Lagrangians, dual to the Fierz-Pauli Lagrangian
(\ref{hlag}) by construction, is the main result of this work.

Let us recall that in (\ref{LAGT1}) the field ${\hat{T}}_{(\mu
\nu )\sigma }$ is only antisymmetric and traceless, therefore
having 20 components, but the Euler-Lagrange equations give rise
to the necessary Lagrangian constraints leading to the final
required five independent degrees of freedom. The final equations
of motion together with the independent Lagrangian constraints
which completely determine the dynamics are
\begin{eqnarray}
\partial _{\alpha }\,\left( \partial ^{\alpha }\hat{T}^{(\nu \rho )\sigma}
+\partial ^{\nu }\hat{T}^{(\rho \alpha )\sigma }+\partial ^{\rho }\hat{T}
^{(\alpha \nu )\sigma }\right) -\frac{1}{4}\left( Q+2P\right) {\hat{T}}
^{(\nu \rho )\sigma }-\frac{1}{2}R\,\epsilon ^{\nu \rho \alpha \beta }
\hat{T}
_{(\alpha \beta )}{}^{\sigma }=0, \nonumber \\
\hat{T}_{(\mu \beta )}{}^{\beta } =0,\quad \epsilon_{\alpha \beta \gamma
\lambda }\,\hat{T}^{(\beta \gamma )\lambda }=0,\quad \partial _{\sigma }
\hat{T}_{(\alpha \beta )}{}^{\sigma }=0,  \nonumber \\
\partial _{\nu }{\hat{T}}^{(\nu \rho )\sigma }+\frac{R}{2P+Q}\,\left(
\epsilon ^{\nu \rho \alpha \beta }\,\partial _{\nu }\hat{T}_{(\alpha \beta
)}{}^{\sigma }+\epsilon ^{\nu \sigma \alpha \beta }\,\partial
_{\nu }\hat{T}
_{(\alpha \beta )}{}^{\rho }\right) =0.  \label{ec4}
\end{eqnarray}

To be sure we have the correct number of degrees of
freedom and to identify the mass parameter it is convenient to
use plane wave solutions in a rest frame, where $k^{\mu }=(\mu
,{\bf 0})$. In this frame the equations of motion read
\begin{equation}
\left( \frac{1}{4}(Q+2P)+\mu ^{2}\right) {\hat{T}}^{(\nu \rho
)\sigma }+\mu ^{2}\,\left( g^{0\nu }\hat{T}^{(\rho 0)\sigma
}+g^{0\rho }\hat{T}^{(0\nu )\sigma }\right)
+\frac{1}{2}R\,\epsilon ^{\nu \rho \alpha \beta }\hat{T}
_{(\alpha \beta )}^{\;\;\;\;\sigma }=0
\end{equation}
with the constraints
\begin{equation}
\hat{T}_{(\alpha \beta )}{}^{0}=0,\quad
{\hat{T}}^{(0i)k}=-\frac{R}{2P+Q} \,\left( \epsilon
^{imn}\,\hat{T}_{(mn)}{}^{k}+\epsilon ^{kmn}\,\hat{T}
_{(mn)}{}^{i}\right) .
\end{equation}
Therefore, all the degrees of freedom can be expressed in terms
of the nine components $\hat{T}_{(ij)k}$, which satisfy the four
constraints
\begin{equation}
\hat{T}_{(ki)}{}^{i}=0,\qquad \epsilon _{ijk}\,\hat{T}^{(ij)k}=0.
\end{equation}
and hence, as expected, five degrees of freedom remain. Using
these constraints the equations of motion become
\begin{equation}
\left( \frac{1}{4}(Q+2P)+4\frac{R^{2}}{Q+2P}+\mu ^{2}\right)
\hat{T} ^{(jk)l}=0,
\end{equation}
which using (\ref{rell}) gives the mass $\mu =M$ for the
${\hat{T}}^{(jk)l}$ field.

The divergence $\partial _{\kappa }T^{(\kappa \lambda )\sigma }$
is non null if $\left( 2m+n-4d\right) \neq 0$. From Eq. (\ref{t2})
the on-shell duality relation can be written
\begin{equation}
T^{(\kappa \lambda )\sigma }=eC^{-1}\left( -\left( 2m+n-4d\right)
\left(
\partial ^{\lambda }h^{\sigma \kappa }-\partial ^{\kappa }h^{\sigma \lambda
}\right) +\left( 2a+c\right) \epsilon _{\mu \nu }^{\;\;\kappa
\lambda }\partial ^{\nu }h^{\sigma \mu }\right) ,  \label{ddual}
\end{equation}
and we see that the divergence is proportional to $h^{\sigma
\lambda }$ on the equations of motion
\begin{equation}
\partial _{\kappa }T^{(\kappa \lambda )\sigma }=-eC^{-1}M^{2}\left(
2m+n-4d\right) h^{\sigma \lambda }\,.  \label{diver}
\end{equation}

The linear combination of the tensors appearing on the right hand
side of (\ref {ddual}) is the most general form that can take the
tensor $T^{(\mu\nu)\sigma}$ of symmetry $ (2,1)$ as a function of
$h^{\sigma \kappa }$ on the equations of motion. Taking $\left(
2m+n-4d\right) =0$, the second term on the right hand side gives
the standard form of the duality involving the Levi-Civita tensor.
For null spatial momentum this becomes $T^{(ij)k}\sim \epsilon
_{l}^{\;ij}h^{lk}$ . This is the duality, in the sense of
representations of $SO(n)$, for the tensors of types $(1,1)$ and
$(2,1)$ in three spatial dimentions. On the other hand the term
proportional to $\left( \partial ^{\lambda }h^{\sigma \kappa
}-\partial ^{\kappa }h^{\sigma \lambda }\right) $ makes the
symmetric part of the divergence $\partial _{\kappa }T^{(\kappa
\mu )\nu }$ non null and proportional to $h_{\mu \nu }$. It
corresponds to giving $T^{(\mu\nu)\sigma}$ some longitudinal
components, as can be seen in the zero spatial momentum where $
T^{(0i)j}\sim h^{ij}$. Thus, the physical degrees of freedom of
$h_{\mu \nu } $ are mapped either on the null divergence part of
$T^{(\mu\nu)\sigma}$ or on the symmetric part of the divergence
$\partial _{\kappa }T^{(\kappa \mu )\nu }$, and in the general
case on a linear combination of both. The case where $h_{\mu \nu
}$ is completely mapped on the divergence of $T^{(\mu\nu)\sigma}$
can not be achieved with the Lagrangian (\ref{plag}) because it
would require $\left( 2a+c\right) =0$, in which case the
transformation becomes ill defined, but can be realized including
the aditional duality term $T_{(\mu \nu )\sigma }\partial ^{\mu
}h^{\nu \sigma }$. We note that although all the parent
Lagrangians with three free parameters represent the same theory,
they give place to different duality mappings, characterized by
one parameter.

We can see that in the parity-conserving case
$2m+n-4d=0\,$($R=0$), the equations of motion and the Lagrangian
constraints simplify to
\begin{equation}
\left( \partial ^{2}+M^{2}\right) {\hat{T}}^{(\nu \rho )\sigma
}=0,\quad { \hat{T}}^{(\nu \rho )}{}_{\rho }=0,\quad \epsilon
_{\kappa \nu \rho \sigma }\,{\hat{T}}^{(\nu \rho )\sigma
}=0,\quad \partial _{\alpha }\,{\hat{T}}^{({ \alpha }\nu )\sigma
}=0,\quad \partial _{\alpha }\,{\hat{T}}^{(\nu \rho )\alpha }=0.
\end{equation}

\section{Massless spin-two dual theories}
Here we will include only a few comments in relation with the
zero-mass case, as a detailed discussion is postponed for a
separate publication. To count the independent degrees of freedom
we can use the Hamiltonian analysis of the zero-mass limit
$(P=Q=R=0)$ of the Lagrangian (\ref{LAGT1}). The $ (3+1)$
splitting of degrees of freedom produces the coordinates
${\hat{T}}
^{i00},\,{\hat{T}}^{0ij},\,{\hat{T}}^{ij0},\,{\hat{T}}^{ijk}$
together with their corresponding canonically conjugated momenta
$\Pi _{i00},\,\Pi _{0ij},\,\Pi _{ij0},\,\Pi _{ijk}$, satisfying
the standard equal-time non-zero Poisson brackets. The whole set
of constraints, which include up to terciary constraints, is
\begin{eqnarray}
&&\Gamma ^{i}=\Pi ^{i00},\quad \Gamma ^{ij}=\Pi ^{0ij},\quad
\Gamma \;=\Pi -4F^{0},\quad \Delta ={\hat{T}}^{0m}{}_{m},\quad
\Delta ^{i}={\hat{T}}
_{\;\;\;\;}^{i00}+{\hat{T}}^{im}{}_{m},  \nonumber \\
&&\Sigma =\epsilon ^{ijk}\partial _{k}\Pi _{ij0},\quad \Sigma
^{ij}=\partial _{m}\Pi ^{mij}-\frac{1}{3}g^{ij}\partial _{m}\pi
^{m},\qquad (\Sigma
^{ij}g_{ij}=0),  \nonumber \\
&&\Phi _{jm}=\partial _{m}\partial ^{q}\Pi _{jq0},\qquad
(\partial ^{j}\Phi _{jm}=0,\quad g^{jm}\Phi _{jm}=0),
\end{eqnarray}
with $\Pi =-\frac{1}{2}\,\epsilon _{ijk}\,\Pi
^{ijk},\,\,F^{0}=-\frac{1}{2} \,\epsilon _{ijk}\,\partial
^{i}\,{\hat{T}}^{jk0},\,\,\pi ^{m}=\Pi ^{mk}{}_{k}$. In
parenthesis we have indicated the identities satisfied by the
corresponding constraints. Due to this we have $31$ independent
constraints. The first class constraints turn out to be the
$9-1=8$ traceless components of $\Gamma ^{ij}$ together with the
$9-3-1=5$ independent components of $\Phi ^{jm}$. Thus, $13$
constraints are first class, while the remaining 18 are second
class. This gives $\frac{1}{2} (2\times 24-2\times 13-18)=2$
independent degrees of freedom in the configuration space.

Finally, a comment on the possible dual systems that could arise
in the massless case. Let us recall the situation in the case of
$p$-forms. Basically there are two ways of taking the zero-mass
limit. One arises directly from the Lagrangian (\ref{FPL}) and
gives
\begin{equation}
{\cal L}_{P}=\frac{1}{2}L\wedge {}^{*}L+L\wedge dB.  \label{M01}
\end{equation}
The second possibility is to rescale the forms $B\rightarrow
M^{-1}$ $B,$ $ L\rightarrow ML\,$ in (\ref{FPL}), which leads to
the alternative massive Lagrangian
\begin{equation}
{\cal L}_{P}=\frac{M^{2}}{2}L\wedge {}^{*}L+L\wedge
dB+\frac{1}{2}B\wedge {}^{*}B,  \label{FPL1}
\end{equation}
together with the corresponding zero mass limit
\begin{equation}
{\cal L}_{P}=L\wedge dB+\frac{1}{2}B\wedge {}^{*}B.  \label{M02}
\end{equation}
In the case of $M\neq 0$ the theories generated by the
Lagrangians (\ref{FPL}) and (\ref{FPL1}) are equivalent.
Nevertheless, for $M=0$ the situation is different. In fact, the
Lagrangian (\ref{M01}) gives the equation of motion $
dL=0\Rightarrow L=dA,$ where $A$ is a $(d-q-2)$-form, thus
producing a duality of the type $q\Leftrightarrow (d-q-2)$. In an
analogous way, the Lagrangian (\ref{M02}) produces the equation
of motion $dB=0$, thus introducing a $(q-1)$-form $C$ such that
$B=dC.$ Here the duality is of the type $(q-1)\Leftrightarrow
(d-q-1).$ Both types of dualities are not equivalent. The two
possible dualities mentioned above arise from the mixing term
$L\wedge d\,B$, plus the choice of the auxiliary field. In our
case, the corresponding mixing term is
\begin{equation}
e\,T_{(\mu \nu )\sigma }\epsilon ^{\mu \nu \alpha \beta }\partial
_{\alpha }h_{\;\beta }^{\sigma },
\end{equation}
which gives the alternative field equations
\begin{equation}
\partial _{\alpha }\left( \epsilon ^{\mu \nu \alpha \beta }T_{(\mu \nu
)}{}^{\sigma }+\epsilon ^{\mu \nu \alpha \sigma }T_{(\mu \nu
)}{}^{\beta }\right) =0,\qquad \epsilon ^{\mu \nu \alpha \beta
}\partial _{\alpha }h_{\sigma \beta }=0.
\end{equation}
Let us emphasize that they do not correspond to a condition of
the form $ dB=0 $, whose solution is given in terms of the
Poincare lemma. Recent generalizations of the Poincare lemma for
mixed symmetry tensors \cite{MH} could be useful to find the
general solutions of the equations above, leading to an explicit
construction of the corresponding dual systems.

{\bf Acknowledgments:} This work was partially supported by
CONICET-Argentina, CONACYT-M\'{e}xico, and Universidad Nacional
Aut\'{o}noma de M\'{e}xico under grant DGAPA- UNAM IN-100397. RM
acknowledges interesting talks with C. Fosco and LFU also
acknowledges useful conversations with A. Garc\'{\i}a.

\end{document}